# Quantum Causal Networks

**Kathryn B. Laskey**
George Mason University
MS 4A6, 4400 University Drive
Fairfax, VA 22030-4400
[klaskey]@gmu.edu

## Abstract

Intervention theories of causality define a relationship as causal if appropriately specified interventions to manipulate a putative cause tend to produce changes in the putative effect. Interventionist causal theories are commonly formalized by using directed graphs to represent causal relationships, local probability models to quantify the relationship between cause and effect, and a special kind of conditioning operator to represent the effects of interventions. Such a formal model represents a family of joint probability distributions, one for each allowable intervention policy. This paper interprets the von Neumann formalization of quantum theory as an interventionist theory of causality, describes its relationship to interventionist theories popular in the artificial intelligence literature, and presents a new family of graphical models that extends causal Bayesian networks to quantum systems.

## 1 INTRODUCTION

Causality is fundamental to artificial intelligence. Intelligent systems must choose actions that are likely to bring about their goals, must monitor situations to detect when plans are going awry, and must modify their plans in appropriate ways in response to unexpected occurrences. Intelligent systems must also learn about causal relationships from interacting with and observing their environments. These capabilities require reasoning about, drawing inferences about, and acting appropriately with respect to, cause and effect relationships in the world.

Causality has been a contentious topic in philosophy and the sciences. The intuitive notion that manipulating causes produces changes in effects has been difficult to formalize as a scientifically rigorous, non-circular theory of causation (cf., Woodward, 2001). Another challenge has been the development of learning methods that can distinguish spurious correlation from genuine causation. In recent years, the artificial intelligence and statistics communities have converged on graphical models as a formal language for expressing cause and effect relationships. Causal graphical models augment graphical probability models with assumptions about the effects of external interventions on the probabilities encoded by the model.

For example, causal Bayesian networks (Pearl, 2000) augment ordinary Bayesian networks with a set of "local surgery" operators that specify how the joint probability distribution behaves with respect to external interventions. The mathematics of local surgery formalizes the idea of causal relationships as stable physical mechanisms, each of which can be perturbed by local interventions without disturbing the others.

Formally, a causal Bayesian network (CBN) for random variables $X_1, \ldots, X_n$ consists of a directed acyclic graph $G$ whose nodes correspond to the $X_i$; a joint distribution $P(\cdot)$ that respects the independence relationships encoded by $G$, and an operator $do(X_i=x_i)$ for each node of the graph that represents an intervention to set $X_i$ to the value $x_i$. The operator $do(X_i=x_i)$ disconnects $X_i$ from its parents and sets its value to $x_i$, while leaving the remainder of the causal relationships and local probability models undisturbed. That is, if $V = (X_{i_1}, \ldots, X_{i_k})$ denotes a subsequence of the $X_i$, and P*$(X_1, \ldots, X_n \mid do(V=v))$ denotes the probability distribution obtained by applying the $do(\cdot)$ operator to set the random variables in $V$ to the values $v = (x_{i_1}, \ldots, x_{i_k})$, then P*$(X_1, \ldots, X_n \mid do(V=v))$ is also a Bayesian network with graph $G$. In this new Bayesian network, the local distributions for variables in $V$ place probability 1 on $V=v$. That is, Pr*$(X_i = x_i \mid pa(X_i)) = 1$ for $i = i_1, \ldots, i_k$, where pa$(X_i)$ denotes the parents of $X_i$ in $G$. The local probability models for the other random variables remain unchanged. That is, P*$(X_i \mid pa(X_i)) = Pr(X_i \mid pa(X_i))$ for $i \neq i_1, \ldots, i_k$. A causal Bayesian network admits two kinds of conditioning: ordinary Bayes conditioning $P(Y|V=v)$ and causal conditioning, i.e., P$(Y \mid do(V=v))$.

Pearl assumes that each cause and effect relationship corresponds to an autonomous physical process involving a relatively small number of variables. A causal Bayesian network describes a collection of such relationships that interact with each other via shared variables. It provides a model to predict the effects of local interventions that change one or more of the variables without disturbing any of the other mechanisms.

Pearl explicitly adheres to the classical view of physical causation, in which all mechanisms are deterministic and probability arises from ignorance of boundary conditions. Although he argues forcefully that a formal mathematics of causation is needed to guard against the errors and confusions of unaided intuition, he argues for classical determinism on the basis of its intuitive appeal:

> … The few esoteric quantum mechanical experiments that conflict with the Laplacian conception evoke surprise and disbelief… Our objective is to preserve, explicate and satisfy – not destroy – those intuitions.

Classical physics has been superseded by the explicitly probabilistic quantum theory. As formalized by von Neumann (1955), quantum theory can be understood as an interventionist causal theory of the kind popularized by Pearl. Extending graphical model based theories of causation into the quantum realm opens the possibility of a theory of cause and effect that is in full agreement with fundamental theory of the physical world.

## 2 QUANTUM THEORY

According to the classical nineteenth century worldview, physical systems followed precisely defined trajectories that evolved according to deterministic laws. Physical theory was causally closed, having no place for interventions into its unfolding. Early in the twentieth century, this classical picture was overturned by a new fundamental physical theory.

Unlike its classical predecessor, quantum theory is stochastic and causally open. Quantum theory represents not only the passive evolution of closed physical systems, but also the effects of interventions. Bohm (1951) said that the quantum state has been called a wave of probability, but it is more accurate to call it a "wave from which many related probabilities can be calculated." In other words, the quantum state predicts not what will occur, nor a single probability distribution for what will occur, but rather a *set* of probability distributions, one for each conceivable intervention that could be made on a quantum system. An intervention results in a stochastic transformation from the state just prior to the intervention to one of the allowable results of the intervention. Quantum theory specifies a probability distribution for the outcome of each such intervention. Thus, quantum theory is naturally viewed as an interventionist theory of causality of the sort that has become popular recently in statistics and the social sciences (Woodward, 2001).

### 2.1 QUANTUM STATES

States of quantum systems are represented mathematically as density operators acting on a Hilbert space $\mathcal{H}$, called the state space of the system. A density operator can be identified with a positive, unit-trace complex-valued square matrix. The dimension of the matrix is a characteristic of $\mathcal{H}$, and can be finite or countably infinite.

States of a composite quantum system are represented as density operators on the tensor product $\mathcal{H}_1 \otimes \cdots \otimes \mathcal{H}_p$ of the Hilbert spaces $\mathcal{H}_1, \mathcal{H}_2, \ldots, \mathcal{H}_p$ for the component systems. Tensor product spaces are the quantum analogue of Cartesian product spaces. A density operator on $\mathcal{H}_1 \otimes \cdots \otimes \mathcal{H}_p$ is called a product state if it can be written as $\sigma_1 \otimes \sigma_2 \otimes \cdots \otimes \sigma_p$, where $\sigma_i$ is a density operator on $\mathcal{H}_i$. States that can be written $\sum_i q_i \sigma_{i1} \otimes \sigma_{i2} \otimes \cdots \otimes \sigma_{ip}$, where the $q_i$ are positive real numbers summing to 1, are called separable states. States that are not separable are called entangled. Entangled quantum systems are responsible for many of the most intriguing and puzzling features of quantum theory. Entangled systems can exhibit long-distance correlations that cannot be explained by local hidden variable theories. Correlations due to entanglement do not reflect causal relationships, in a sense to be made precise below.

Given a quantum state $\sigma$ on a tensor product space $\mathcal{H}_1 \otimes \cdots \otimes \mathcal{H}_p$, a reduced density operator $\sigma_i$ on the $i^{\text{th}}$ Hilbert space can be obtained via an operation called the partial trace. The reduced density operator is the quantum analogue of the marginal distribution for a classical joint distribution. The reduced density operator correctly describes the statistical behavior of observable quantities of a subsystem, when attention is restricted to quantities pertaining to the given subsystem.

Cerf and Adami (1999) propose a quantum analogue for the classical conditional distribution. Warmuth and Kuzmin (2006) extend Cerf Adami's conditional density operators to propose a generalization of the Bayesian probability calculus to density matrices. These authors do not address causality.

## 2.2 EVOLUTION OF QUANTUM SYSTEMS

Traditional quantum theory as formalized by von Neumann (1955) specifies two kinds of transformations quantum systems can undergo. One of these is continuous and reversible evolution of systems isolated from environmental effects. If the initial state for an isolated quantum system is the density operator $\sigma(t_0)$, then at time $t_1 > t_0$ the state will be:

$$\sigma(t_1) = U(t_1-t_0) \, \sigma(t_0) \, U(t_1-t_0)^*, \qquad (1)$$

where $U(t)$ is a unitary operator[1] given by:

$$U(t) = \exp\{-iHt/\hbar\}, \qquad (2)$$

$H$ is a Hermitian (i.e., self-adjoint) operator on $\mathcal{H}$ called the *Hamiltonian*, and $\hbar$ is Planck's constant divided by $2\pi$.

The other kind of transformation is a stochastic state change that has been called state reduction, projective measurement, or more picturesquely, collapse. The term reduction is preferred here because it is more neutral than collapse, and emphasizes that stochastic transformations apply to a broader class of problems than measurements performed in laboratory experiments. Reduction is represented mathematically as a discontinuous transformation at time $t$ from the state $\sigma(t-)$ to the state $\sigma(t+)$. With the reduction event is associated a set $\{P_i\}$ of mutually orthogonal projection operators on $\mathcal{H}$ that sum to the identity, i.e.:

i. $P_i^2 = P_i$;
ii. $P_i P_j = 0$ for $i \neq j$; and
iii. $\sum_i P_i = I$.

The possible outcomes of the reduction are density operators $P_i \sigma(t-) P_i / \mathrm{Tr}(\sigma(t-))$, where $\mathrm{Tr}(\cdot)$ denotes the trace operator, or sum of diagonal elements of the matrix. Division by $\mathrm{Tr}(\sigma(t-))$, called *normalization*, preserves the unit trace property of density operators. Conditional on the time $t$ at which the reduction occurs and the set $\{P_i\}$ of projection operators, the outcome probabilities are given by the Born rule: the probability that the outcome of applying the projection set $\{P_i\}$ when the system is in state $\sigma(t-)$ is given by:

$$\mathrm{Tr}(P_i \sigma(t-) P_i) / \mathrm{Tr}(\sigma(t-)). \qquad (3)$$

Because there are at most $n$ mutually orthogonal projection operators of dimension $n$, the number of possible outcomes of any reduction can be no more than the dimension of the system's Hilbert space. Thus, a density operator on an $n$ dimensional Hilbert space is the quantum analogue of a probability distribution for a random variable with $n$ possible outcomes. Whereas a classical random variable represents outcome probabilities for a single experiment with a given set of $n$ possible outcomes, a density operator represents outcome probabilities for an infinite collection of experiments, each with a different set of $n$ possible outcomes. In particular, quantum probabilities are contingent: *if* the experiment associated with the set $\{P_i\}$ is carried out on a system in state $\sigma(t-)$, *then* the outcome probabilities are given by Equation (3). On the other hand, if a different experiment is carried out, then the outcome probabilities are the Born probabilities for that experiment.

Quantum theory as thus formulated is an explicitly temporal theory. Unitary evolution proceeds from past to future, and reductions are instantaneous discontinuous state changes. In relativistic physics, the temporal ordering of two events may depend on the frame of reference. Quantum theory as described in this section is consistent with relativity theory if it is assumed that reductions occur along spacelike surfaces (cf. Stapp, 2001).

## 2.3 QUANTUM ONTOLOGY

Although the empirical predictions of quantum theory have been confirmed to a high degree of accuracy, the ontological status of quantum probabilities and reductions are a matter of intense controversy. Many scientists subscribe to Einstein's view that "God does not play dice," and are reluctant to embrace an intrinsically stochastic theory as fundamental. Many are also uncomfortable with the incompleteness of quantum theory. Quantum theory makes predictions conditional on interventions, but has nothing to say about the laws governing the occurrence of interventions. The founders of quantum theory assigned interventions to the free choice of human observers, and placed that choice outside of quantum theory. There have been many attempts to found quantum theory on unitary evolution alone, and to explain reductions as artifacts of applying the partial trace operator to restrict attention to subsystems entangled with their environments. Although these theories have passionate advocates, controversy remains over whether dispensing with reductions is possible.

---

[1] The operator $U$ is unitary if its inverse $U^{-1}$ is equal to its adjoint $U^*$.

Because there is no question that von Neumann theory is in accord with observation, and because it provides a natural quantum analogue to classical causal Bayesian networks, this paper adopts the von Neumann formulation. Just as with causal Bayesian networks, the mathematical formalism of quantum causal networks can be adopted as a pragmatic computational tool regardless of one's metaphysical position regarding the ontological status of reductions. A deeper debate on the ontology of quantum theory is beyond the scope of this paper.

## 2.4   QUANTUM OPERATIONS

Quantum operations provide a unified representation for transformations of isolated systems, systems that interact with their environments, and reductions. The quantum operations formalism is equivalent to the von Neumann formalism described above, in that any quantum operation can be represented as a composition of unitary operators, stochastic projections, and partial traces (Nielsen and Chuang, 2000). Because of their generality, quantum operations are seeing wide application to analyzing the behavior of quantum systems, especially in quantum computing and quantum information theory.

Quantum operations are especially useful building blocks for a theory of quantum causality, because they can describe quantum transformations in which the input and output systems are different. That is, quantum operations can represent interactions in which the behavior of one system causes changes in the state of a second system, without requiring an explicit representation of the prior state of the affected system or the post-interaction state of the system producing the effect.

A quantum operation $\mathcal{A}(\sigma)$ is a linear map that transforms operators on an *input* Hilbert space to operators on an *output* Hilbert space, such that the following conditions are satisfied:

1. $\text{Tr}(\mathcal{A}(\sigma)) \leq \text{Tr}(\sigma)$;
2. $\mathcal{A}(\cdot)$ is a *completely positive* map. That is, if $\sigma$ is a positive operator on the input space, then $\mathcal{A}(\sigma)$ is a positive operator on the output space. Furthermore, if $n$ is a positive integer, $\rho$ is a positive operator on the tensor product of an auxilliary $n$-dimensional Hilbert space and the input space, and $\mathcal{I}_p$ is the identity operator on the auxiliary space, then $(\mathcal{I}_p \otimes \mathcal{A})(\rho)$ is a positive operator.

The partial trace operation that maps a density operator for a composite system to the reduced density operator for a subsystem is an example of a quantum operation. Unitary transformations are also quantum operations. If $P$ is a projection operator, the map from $\sigma$ to $P\sigma P$ is a quantum operation that does not preserve the trace.

To satisfy the requirements of relativity theory, if a quantum operation represents temporal evolution, the output system must be localized within the future light cone of the input system. In the case of stochastic reduction operations or partial trace operations in which the output represents a subsystem of a composite system, the output system must not overlap the past light cone of the input system.

## 2.5   PROPER AND IMPROPER MIXED STATES

A density operator is called a pure state if it has rank one; otherwise, it is called a mixed state. If $\sigma$ is a rank $k$ density operator on a Hilbert space $\mathcal{H}$, then there exist pure states $\sigma_1, \ldots, \sigma_k$, and positive real numbers $p_1, \ldots, p_k$, such that:

$$\sum_i p_i = 1 \text{ and } \sum_i p_i \sigma_i = \sigma. \tag{4}$$

Because of this weighted sum representation, mixed states have been interpreted as representing uncertainty about the state of a system. That is, the mixed state $\sum_i p_i \sigma_i$ can represent a system that has probability $p_i$ of being in pure state $\sigma_i$. This decomposition as a weighted sum of pure states may not be unique. A state $\sigma = \sum_i p_i \sigma_i = \sum_i r_i \rho_i$ with two different decompositions as probability-weighted sums of pure states could represent either a system having probability $p_i$ of being in state $\sigma_i$, or a system having probability $r_i$ of being in state $\rho_i$.

When there is entanglement, the reduced density operator of a subsystem may be in a mixed state even when the composite system is in a pure state. Such a subsystem cannot be said to possess a definite state of its own, independent of its environment. Mixed states reflecting uncertainty about definite pure states are called proper mixtures; mixed states arising from entanglement are called improper mixtures (d'Espagnat, 1976).

For subscribers to an ontology that dispenses with reductions, all mixtures are improper. Nevertheless, there is a useful distinction between systems that behave as proper or improper mixtures relative to a given experimental context (Timpson and Brown, 2005). Thus, statements about proper and improper mixtures can be interpreted, with appropriate qualifications regarding the experimental context, within a no-reduction ontology. Proper mixtures can be empirically distinguished from improper mixtures if the system and its environment can be observed jointly, but cannot be empirically distinguished if observations are restricted to the system in isolation.

In the formalism developed here, improper mixtures are represented as mixed states, and proper mixtures are represented as probability distributions over the mixture components. Trace-preserving quantum operations are interpreted as transitions that do not have multiple physically distinguishable outcomes. If outcomes of a transition are distinguishable but unknown, then its result is represented as a probability distribution over the distinguishable outcomes.

Trace-reducing quantum operations represent stochastic transformations with physically distinguishable outcomes. Consider a set $\mathcal{A}_1(\cdot), \ldots, \mathcal{A}_n(\cdot)$ of trace-reducing quantum operations such that $\sum_i \text{Tr}(\mathcal{A}_i(\sigma)) = \text{Tr}(\sigma)$ for all $\sigma$. This set represents a situation in which a transformation is chosen by a stochastic rule. The probability that the $i^{\text{th}}$ transformation occurs is given by $\text{Tr}(\mathcal{A}_i(\sigma))$, and the result of the $i^{\text{th}}$ transformation on input $\sigma$ is $\mathcal{A}_i(\sigma)/\text{Tr}(\sigma)$. We make an explicit distinction between the proper mixture representing ignorance of the result of applying the set of trace-reducing operations $\mathcal{A}_1(\cdot), \ldots, \mathcal{A}_n(\cdot)$, and the improper mixture that results from applying the trace-preserving operation $\sum_i \mathcal{A}_i(\cdot)$. The result of the former is represented in our formalism as a probability distribution over states; the result of the latter is represented as a mixed state.

## 2.6 FIDUCIAL PROJECTIONS

When the state space has dimension $n$, there exists a set $F_1, \ldots, F_{n^2}$ of projection operators, such that the state is characterized by the Born probabilities associated with the $F_i$ (Nielson and Chuang, 2000). Any such collection $\{F_i\}$ is called a set of *fiducial* projections (Hardy, 2001). If $\{F_i\}$ is a fiducial set, and $\sigma$ and $\rho$ are two density operators such that $\text{Tr}(F_i\sigma F_i) = \text{Tr}(F_i\rho F_i)$ for $i = 1, \ldots, n^2$, then $\sigma = \rho$. The fiducial projections can be chosen to have rank 1. In this case, the fiducial projections are themselves density operators, and they represent pure states of the system. Because $F_i$ is a projection operator with rank 1, it can be shown that if $F_i\sigma F_i \neq 0$, then $F_i\sigma F_i/\text{Tr}(F_i\sigma F_i) = F_i$.

A fiducial projection operator $F_i$ thus represents both a pure state of the system and an intervention that has $F_i$ as one of its possible outcomes. If the intervention $F_i$ is applied to a system whose pre-intervention state is $\sigma$, then the probability is $\text{Tr}(F_i\sigma F_i)$ that the post-intervention state is to $F_i$. Because of noncontextuality, these probabilities hold for *any* intervention in which $F_i$ is one of the possible outcomes, regardless of the other possible outcomes of the intervention.

Just as quantum states can be characterized by the probabilities associated with fiducial operators, quantum operations can be characterized by how they act on fiducial operators. Specifically, let $F_1, \ldots, F_{n^2}$ be a set of fiducial projectors on an $n$-dimensional input Hilbert space and let $G_1, \ldots, G_{m^2}$ be a set of fiducial projectors on an $m$-dimensional output space. Suppose that $\mathcal{A}(\cdot)$ and $\mathcal{A}'(\cdot)$ are completely positive maps such that $\text{Tr}(G_j\mathcal{A}(F_i)G_j) = \text{Tr}(G_j\mathcal{A}'(F_i)G_j)$ for $i=1,\ldots,n$ and $j=1,\ldots,m$. Then $\mathcal{A}(\cdot)$ is equal to $\mathcal{A}'(\cdot)$ (Nielsen and Chuang, 2000, sec. 8.4.2).

# 3 QUANTUM CAUSAL NETWORKS

Quantum causal networks formalize cause and effect relationships in quantum systems. In a QCN, a graph represents dependence relationships, quantum operations represent quantitative information about state evolution, and intervention operators represent reductions. QCNs differ from Tucci's (1995) causal Bayesian networks (QBNs), in that QCNs formalize cause and effect relationships in terms of effects of interventions, whereas Tucci's quantum Bayesian networks generalize non-causal Bayesian networks to quantum systems.

## 3.1 SEQUENCED ASSOCIATION GRAPHS

In causal Bayesian networks, the arcs are directed and the probabilistic dependencies are causal. Of course, it is easy to find real-world examples of correlations that do not correspond to causal relationships.

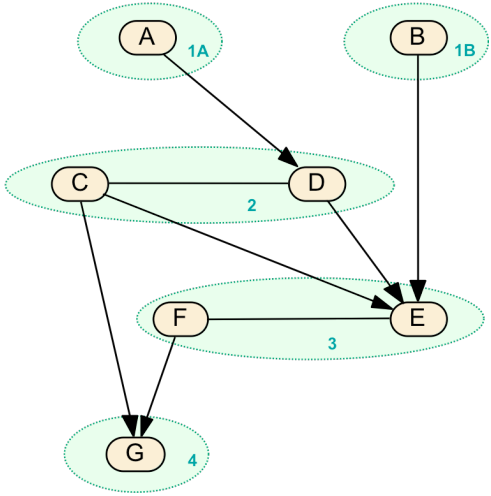

**Figure 1: Sequenced Association Graph**

Nevertheless, outside the quantum realm, it is generally assumed that Riechenbach's principle of common causes holds. That is, when two quantities are correlated, it is assumed either that one is a cause of the other or that there is another variable that is a common cause of both. When the principle of common cause holds, one can construct a causal Bayesian network by inserting variables as needed to represent common causes of correlated variables.

In quantum systems, although entanglement can give rise to correlations between spacelike separated events, causal influence can operate only between timelike separated events, and only from past to future. This fundamental difference between correlations involving spacelike and timelike separated events is represented in sequenced association graphs by using directed arcs to represent causal influences from the past to the future, and undirected arcs to represent contemporaneous correlations between entangled systems.

***Definition* 1:** Let $G$ be a graph, and let $A$ and $B$ be nodes of $A$. Then $A$ and $B$ are *contemporaneous* if (*i*) there is an undirected edge connecting $A$ and $B$, or (*ii*) there is an undirected edge between $A$ and a node contemporaneous with $B$. If $A$ and $B$ are contemporaneous, we write $A \sim_T B$.

***Definition* 2:** Let $G$ be a graph, and let $A$ and $B$ be nodes of $A$. Then $A$ *precedes* $B$ if (*i*) there is a directed edge from $A$ to $B$, or (*ii*) there is a directed edge from $A$ to a node that precedes $B$. If $A$ precedes $B$, we write $A \prec_T B$.

A straightforward inductive argument shows that $\sim_T$ is an equivalence relation and $\prec_T$ is transitive.

***Definition* 3:** A graph $G$ is a *sequenced association graph* (SAG) if there is no pair of nodes $A$ and $B$ such that (*i*) $A$ precedes $B$ and (*ii*) $B$ precedes or is contemporaneous with $A$.

The directed arcs in a sequenced association graph establish a partial order on the nodes. When a SAG is used to model a physical process, each node is associated with a physical system localized within a region of spacetime. Directed edges connect timelike separated systems, and are oriented from past to future. Undirected edges connect spacelike separated systems that are correlated due to entanglement.

Because contemporaneity is an equivalence relation, it partitions the nodes of a SAG into equivalence sets. The elements of this partition are called CN-sets.

***Definition* 4:** Let $G$ be a sequenced association graph. A *CN-set* is a maximal subset of mutually contemporaneous nodes of $G$. A *root CN-set* is a CN-set in which none of the arcs in $G$ enters any of the nodes in the CN-set. A CN-set that is not a root CN-set is called a *child CN-set*.

Figure 1 shows a SAG containing five CN-sets, enclosed in dotted ovals and numbered $1A$ through $4$. The numbering scheme indicates the time order if it can be established from the graph. Letters are appended to the numbers to label nodes for which the order cannot be distinguished. The time ordering of CN-sets $1A$ and $1B$ cannot be determined from the graph; the CN-sets $2$ through $4$ follow these sets in temporal order.

### 3.2 QUANTUM CAUSAL NETWORKS

A quantum causal network (QCN) is a quantum analogue to a CBN. Like a Bayesian network, a QCN uses a graph to represent qualitative relationships and local probability models to represent numerical likelihood information. The graph for a QCN is a sequenced association graph. Numerical likelihood information is represented with density operators and quantum operations.

***Definition* 5:** Let $G$ be a SAG, and let $\{X_1, \ldots, X_k\}$ be a child CN-set for $G$. A node $Y$ is an *influencing parent* for the CN-set if $G$ has a directed edge from $Y$ to one of the $X_i$, and a *non-influencing parent* for the CN-set if it is contemporaneous to a parent for the CN-set.

***Definition* 6:** Let $G$ be a SAG, let $\{X_1, \ldots, X_k\}$ be a CN-set for $G$, and let $\mathcal{H}_i$ denote the Hilbert space associated with $X_i$. Let $\{W_1, \ldots, W_r\}$ denote the set of influencing and non-influencing parents for $\{X_1, \ldots, X_k\}$, and let $\mathcal{F}_i$ denote the Hilbert space associated with $W_i$. A *local distribution* $\Delta(\cdot)$ for $\{X_1, \ldots, X_k\}$ is defined as:

1. If $\{X_1, \ldots, X_k\}$ is a root CN-set, then $\Delta(X_1, \ldots, X_k)$ consists of a finite or countably infinite set $\{\Delta_i(X_1, \ldots, X_k)\}$ of density operators on $\mathcal{H}_1 \otimes \cdots \otimes \mathcal{H}_k$ such that $\Sigma_i \Delta_i(X_1, \ldots, X_k)$ is equal to the identity;

2. If $\{X_1, \ldots, X_k\}$ is a child CN-set, then $\Delta(X_1, \ldots, X_k \mid W_1, \ldots, W_r)$ consists of a finite or countably infinite set $\{\Delta_i(X_1, \ldots, X_k \mid W_1, \ldots, W_r)\}$ of quantum operations mapping $\mathcal{F}_1 \otimes \cdots \otimes \mathcal{F}_r$ to $\mathcal{H}_1 \otimes \cdots \otimes \mathcal{H}_k$, such that $\Sigma_i \mathrm{Tr}(\Delta_i(X_1, \ldots, X_k \mid W_1, \ldots, W_r))$ is trace-preserving.

***Definition* 7:** Let $\mathcal{H}_1 \otimes \cdots \otimes \mathcal{H}_n$ be a product space. A *fiducial reduction* is a set of projection operators satisfying conditions *i-iii*, in which each projector in the set is a product $F_1 \otimes \cdots \otimes F_n$ of fiducial projectors.

***Definition* 8:** Let $G$ be a SAG, and let $\{X_1, \ldots, X_k\}$ be a root CN-set. The local distribution $\Delta(X_1, \ldots, X_k)$ *respects* $G$ if for any fiducial reduction applied to $\Delta(X_1, \ldots, X_k)$ and any $i$, the conditional probability of $X_i$ given $X_1, \ldots, X_{i-1}, X_{i+1}, \ldots, X_k$ depends only on the neighbors of $X_i$ in $G$.

***Definition* 9:** Let $G$ be a SAG, let $\{X_1, \ldots, X_k\}$ be a child CN-set, and let $\{W_1, \ldots, W_r\}$ denote its influencing and non-influencing parents. The local distribution $\Delta(X_1, \ldots, X_k \mid W_1, \ldots, W_r)$ *respects* $G$ if the following condition holds. For $i=1, \ldots, r$, let $F_i$ denote a fiducial projector on the Hilbert space for $W_i$. Let $\Delta(X_1, \ldots, X_k \mid W_1, \ldots, W_r)(F_1 \otimes \cdots \otimes F_r)$ denote the quantum operation $\Delta(X_1, \ldots, X_k \mid W_1, \ldots, W_r)$ applied to the product projector $F_1 \otimes \cdots \otimes F_r$. Then the conditional probability assigned by $\Delta(X_1, \ldots, X_k \mid W_1, \ldots, W_r)(F_1 \otimes \cdots \otimes F_r)$ to $X_i$ given $X_1, \ldots, X_{i-1}, X_{i+1}, \ldots, X_k$ depends only on those $X_j$ that are neighbors of $X_i$ in $G$ and those $W_j$ that are parents of $X_i$ in $G$.

***Definition* 10:** Let $G$ be a sequenced association graph. Let $\{\mathcal{H}_i\}$ be a collection of Hilbert spaces, one for each node $X_i$ of $G$. Let $\{\mathrm{Pr}(\cdot)\}$ be a set of local distributions, one for each CN-set of $G$. Then $Q = (G, \{\mathcal{H}_i\}, \{\mathrm{Pr}(\cdot)\})$ is a *quantum causal network* if each of the local distributions respects $G$.

The density operator for a root CN-set of a QCN requires at most $n^2-1$ real numbers to specify, where $n$ is the product of the dimensions of the Hilbert spaces for the nodes in the CN-set. The quantum operation for a child CN-set requires at most $n^2(m^2-1)$ real numbers, where $m$ is the product of the dimensions of the Hilbert spaces for the parent nodes and $n$ is the product of the dimension of the nodes in the child CN-set. The independence assumptions encoded in $G$ reduce the number of parameters needed to specify these local distributions.

## 3.3 THE JOINT DISTRIBUTION FOR A QCN

A QCN induces a joint density operator on the tensor product of the Hilbert spaces associated with the nodes of its SAG. This joint density operator can be constructed by propagating the quantum operations forward in the direction of the causal arcs. This propagation process induces a density operator on each CN-set. For brevity, we sketch the construction for a two-node QCN with graph $X \rightarrow Y$. The extension to the general case is straightforward.

First, consider the root node $X$ local distribution $\{\Delta_i(X)\}$ for $X$. We define a reduced density operator on $\mathcal{H}_X$ as $\sigma_X = \Sigma_i \Delta_i(X)$. Any density operator can be expanded as a mixture of mutually orthogonal 1-dimensional density operators (cf., Nielsen and Chuang, 2000). Thus, we can write

$$\sigma_X = \sum_i \theta_i Q_i, \quad (5)$$

where the $Q_i$ are mutually orthogonal one-dimensional projection operators on $\mathcal{H}_X$, and the $\theta_i$ are non-negative numbers that sum to 1.

Now consider the local distribution $\{\Delta_i(Y \mid X)\}$ for $Y$. We define the trace-preserving quantum operation $\mathcal{A}(Y \mid X) = \Sigma_i \Delta_i(Y \mid X)$. For each $i$, we can write

$$\mathcal{A}_{Y \mid X}(Q_i) = \sum_j \rho_{ij} R_{ij}, \quad (6)$$

where the $R_i$ are mutually orthogonal one-dimensional projection operators on $\mathcal{H}_Y$, and the $\rho_i$ are non-negative numbers that sum to 1. Note that the mixture components $R_{i1}, R_{i2}, \ldots$ for $\mathcal{A}_{Y \mid X}(Q_i)$ may be different for different $i$.

Next, we form a joint density operator on $\mathcal{H}_X \otimes \mathcal{H}_Y$ as follows:

$$\tau_{XY} = \sum_{i,j} \theta_i \rho_{ij} Q_i \otimes R_{ij}. \tag{7}$$

The density operator $\tau_{XY}$ represents a quantum state for the undisturbed two-node QCN. Applying the partial trace yields density operators $\sigma_X$ and $\sigma_Y$ to represent the states of the $X$ and $Y$ subsystems of the undisturbed joint system.

This construction generalizes in a straightforward manner to construct a density operator on the tensor product space $\mathcal{H}_1 \otimes \cdots \otimes \mathcal{H}_n$. A reduced density operator for each node can be obtained via the partial trace operation. These density operators represent undisturbed evolution of the quantum system. Furthermore, it is straightforward to show that there is a representation of this joint density operator as a mixture of density operators, in which the joint distribution represented by the mixture weights is represented by a graphical model with graph $G$ and local distributions given as follows:

1. If $\{X_1, \ldots, X_k\}$ is a root CN-set, then $(X_1, \ldots, X_k)$ has value $\Delta_i(\xi_1, \ldots, \xi_k)/\text{Tr}(\Delta_i(\xi_1, \ldots, \xi_k))$ with probability $\text{Tr}(\Delta_i(\xi_1, \ldots, \xi_k))$;

2. If $\{X_1, \ldots, X_k\}$ is a child CN-set, then conditional on the state $\omega$ of $(W_1, \ldots, W_r)$, the CN-set $(X_1, \ldots, X_k)$ has value $\Delta_i(\xi_1, \ldots, \xi_k \mid \omega_1, \ldots, \omega_r)/\text{Tr}(\Delta_i(\xi_1, \ldots, \xi_k \mid \omega_1, \ldots, \omega_r))$ with probability $\text{Tr}(\Delta_i(\xi_1, \ldots, \xi_k \mid \omega_1, \ldots, \omega_r))$.

## 4 EFFECTS OF INTERVENTIONS

Just as in causal Bayesian networks, a QCN encodes a model not just for undisturbed evolution of a system, but also for its response to interventions. In our discussion of interventions to a QCN, we consider two kinds of interventions. The first type of intervention is a reduction operator $rd(X \rightarrow \{P_i\})$ corresponding to a stochastic transformation of the state into its projection onto the subspace associated with one of the $P_i$. The second type of intervention is a $do(X=\xi)$ operator like the one for causal Bayesian networks.

Only the first type of intervention is covered by the mathematical rules of quantum theory. Furthermore, its ontological status is the subject of heated debate. Nevertheless, in practical applications and informal discussions of quantum experiments, it is common to postulate that a density operator has been set to a known state by an experimenter. In other words, $do(X=\xi)$ operations are important in informal discourse and pragmatic applications of quantum theory. This situation is very like the situation in statistics, in which there is an extensive and well-developed formal mathematics of joint probability distributions, but causality has until recently been treated informally and pragmatically. Just as the recently developed formal mathematics of causality has led to new insights and methodologies in the realm of classical systems, extensions of this research to the quantum domain may lead to new insights about causality in quantum systems.

We begin our treatment of interventions by considering the application of the reduction operator $rd(X \rightarrow \{P_i\})$ to a deterministic QCN $Q$, in which every local distribution has a single component density operator (for a root CN-set) or quantum operation (for a non-root CN-set). An intervention at a single target node $T$ of $Q$ is modeled as follows. Let CN($T$) denote the CN-set of $T$. Let $\mathcal{H}_{\text{CN}(T)}$ denote the associated Hilbert space, and let $\sigma_{\text{CN}(T)}$ denote the reduced density operator for CN($T$). Let $P_1, \ldots, P_n$ be a set of one-dimensional[2] orthonormal projection operators (i.e., satisfying Conditions i-iii above) on $\mathcal{H}_{\text{CN}(T)}$, such that each $P_i$ acts as the identity on all nodes except $T$. The intervention results in a new QCN $Q^*$, where:

1. The graph $G^*$ of $Q^*$ is obtained from graph $G$ of $Q$ by removing all directed arcs entering CN($T$) and all undirected arcs between $T$ and members of its CN-set.

2. The new local distribution $\Delta^*(\text{CN}(T))$ is the set $\{P_i \sigma_{\text{CN}(T)} P_i\}$. That is, the possible values for CN($T$) are $\sigma_i = P_i \sigma_{\text{CN}(T)} P_i / \text{Tr}(P_i \sigma_{\text{CN}(T)} P_i)$, for $i = 1, \ldots, n$. The probability of obtaining $\sigma_i$ is $\text{Tr}(P_i \sigma_{\text{CN}(T)} P_i)$. Note that all independence relationships among nodes in CN($T$) that existed in $\sigma_{\text{CN}(T)}$ are preserved in $\sigma_i$. Therefore, $\sigma_i$ respects $G^*$.

3. The local distributions for all nodes not in CN($T$) are unchanged.

---

[2] If the projection operators have dimension greater than one, then Rule 1 must be modified to account for the possibility that entanglement with the node's CN-set may not be destroyed.

As a result of the intervention, the target node takes on one of the allowable results for the projection set associated with the reduction operation. If the target node is entangled with contemporaneous neighbors, intervening at the node may affect these neighbors even though the projector acts as the identity on these nodes. Post-intervention states for descendants of the target node's CN-set are obtained by forward propagation. For this reason, when interventions correspond to non-commuting operators, one must specify an order in which they are to be applied.

To extend the discussion to interventions in proper mixture QCNs, we note that proper and improper mixtures behave identically with respect to forward inference, but differ in the treatment of backward inference from effects to causes. Specifically, when a reduction is applied to an improper mixed state, the result is to disconnect the node from its parents and replace its value stochastically with one of the possible outcomes. No backward inference to the parent is licensed, except to note a contradiction if a result is observed whose Born probability is zero. The situation is different with a proper mixture. With a proper mixture, the result of a reduction at a child node provides information about which mixture component obtains at the parent node.

These informal ideas are formalized by the following rules for interventions at a target node $T$ in a general QCN $Q$. Again, let CN($T$) denote the CN-set of $T$. Let $\mathcal{H}_{CN(T)}$ denote the associated Hilbert space, and let $\sigma_{CN(T)}$ denote the reduced density operator for CN($T$). Let $P_1, ..., P_n$ be a set of one-dimensional orthonormal projection operators (i.e., satisfying Conditions *i-iii* above) on $\mathcal{H}_{CN(T)}$, such that each $P_i$ acts as the identity on all nodes except $T$. The intervention results in a new QCN $Q^*$, where:

1. The graph $G^*$ of $Q^*$ is obtained from graph $G$ of $Q$ by removing any directed arcs entering CN($T$) from parents with only a single possible value, as well as any undirected arcs between $T$ and members of its CN-set that have only a single possible value.

2. The new local distribution $\Delta^*(CN(T))$ consists of the transformations $\{P_i \tau_m P_i\}$, where $\tau_m$ is a joint density operator for CN($T$), and $\tau_m$ has non-zero weight under the undisturbed distribution. That is, the possible values for CN($T$) are $\tau_{im} = P_i \tau_m P_i / \text{Tr}(P_i \tau_m P_i)$, for $i = 1, ..., n$, and $\tau_m$ as above. The probability of obtaining $\tau_{im}$ given the state of the parent CN-set is obtained via the Born probability rule. Note that all independence relationships among nodes in CN($T$) that existed in the undisturbed distribution are preserved in $\tau_{im}$. Therefore, $\tau_{im}$ respects $G^*$.

3. The local distributions for all nodes not in CN($T$) are unchanged.

In a mixture QCN, applying a reduction operator to a node $X$ does not disconnect $X$ from its parents.

We now consider how to represent a local surgery operator $do(X=\xi)$, corresponding to intervening and setting $X$ to the value $\xi$. As noted above, this kind of intervention is often employed informally in descriptions of the set-up for quantum mechanical experiments, but there is no physical theory for how such interventions are effected. Our representation must satisfy two criteria. First, we want the intervention to disconnect $X$ from its parents, even in the case of a proper mixture. That is, the process of setting $X$ to the value $\xi$ should occur regardless of the unknown value of pa($X$). Second, the intervention must not have a causal impact on variables entangled with $X$. To accomplish this, the operator $do(X=\xi)$ acts as follows. First, we apply a reduction operator that projects $X$ to a proper mixture with the same mixture weights as its reduced density operator in the undisturbed model, and acts as the identity on the other nodes in CN($X$). This results in a new QCN in which there are no undirected arcs involving $X$. Next, we remove all arcs entering $X$, set the local distribution of $X$ to place probability 1 on the value $\xi$, and leave all other mechanisms in the QCN undisturbed. This procedure has the desired properties.

## 5 DISCUSSION

The predictions of quantum theory have been subjected to extensive empirical testing for a wide variety of quantum processes, with strong agreement between theory and empirical results. However, quantum theory as presently formulated contains a major explanatory gap, having nothing to say about when a reduction will occur and which set of orthogonal projection operators corresponds to the possible results. Despite intense effort over many years, no one has yet found a satisfactory way to dispense with reductions and still bring quantum theory into concordance with the results of measurements, and physicists disagree strongly about the feasibility of the endeavor. Because reductions are associated with scientists performing measurements, the lack of a theory for state reduction has been called the "measurement problem."

Rather than attempt to dispense with reductions, we formalize reductions as external interventions in a causal graphical model formulation of quantum theory. The formalism is consistent with standard von Neumann quantum theory, but is explicated in a language that ties it to recent work on probabilistic models of causality. As with von Neumann quantum theory, the mathematical formalism can be employed independent of one's metaphysical stance on reductions. It is hoped that formulating a quantum version of causal graphical models will shed light on the physical realizability of causal theories.

In particular, Pearl's *do*-calculus can be viewed as a classical approximation to a physically accurate theory of causation. One role for a quantum theory of causality is to explicate conditions under which such an approximation is adequate. A Pearl-style causal Bayesian network is a reasonable approximation when: (1) decoherence effects nearly eliminate the off-diagonal elements of the density operator for all subsystems under consideration, rendering the global system essentially equivalent for all practical purposes to a statistical mixture of quasi-classical states; and (2) it is physically possible to set the state of a subsystem to a desired state without major disturbance to other subsystems. The first condition holds in many cases of practical interest, but at present our understanding of the second condition is mainly heuristic and empirical.

A physically realistic quantum theory of causation may open up new avenues of investigation. Specifically, it opens the door to new, theoretically well-founded research into the kinds of interventions that are physically achievable and the conditions under which they can be applied without disturbing the states of and causal interactions among subsystems other than the targets of intervention.

## Acknowledgments

Appreciation is due to Henry Stapp for many helpful and stimulating discussions, and for sharing his deep knowledge of quantum theory.